\documentclass[twocolumn,superscriptaddress,aps,prl,floatfix]{revtex4-1}
\usepackage{array}
\usepackage{amssymb}
\usepackage{amsmath}
\usepackage{natbib}
\usepackage{graphicx}
\usepackage{multirow}
\usepackage{color}

\begin{document}

\title{Nuclear quantum effects in ab initio dynamics: \\theory and experiments for lithium imide}
\author{Michele Ceriotti} 
\email{michele.ceriotti@phys.chem.ethz.ch}
\affiliation{Computational Science, DCHAB, ETH Zurich, USI Campus, via G. Buffi 13, CH-6900 Lugano, Switzerland}

\author{Giacomo Miceli}
\affiliation{Dept. of Materials Science, Universit\`{a} di Milano-Bicocca, via R. Cozzi 53, I-20125 Milano, Italy}
\affiliation{Computational Science, DCHAB, ETH Zurich, USI Campus, via G. Buffi 13, CH-6900 Lugano, Switzerland}

\author{Antonino Pietropaolo} 
\affiliation{CNISM UdR Roma Tor Vergata and Centro NAST Universit\`{a} degli Studi di Roma Tor Vergata, via della Ricerca Scientifica 1, I-0133 Roma, Italy}

\author{Daniele Colognesi}
\affiliation{Istituto dei Sistemi Complessi, CNR, via Madonna del Piano 10, 50019 Firenze, Italy}

\author{Angeloclaudio Nale} 
\affiliation{Dept. of Materials Science, Universit\`{a} di Milano-Bicocca, via R. Cozzi 53, I-20125 Milano, Italy}

\author{Michele Catti}
\affiliation{Dept. of Materials Science, Universit\`{a} di Milano-Bicocca, via R. Cozzi 53, I-20125 Milano, Italy}

\author{Marco Bernasconi} 
\affiliation{Dept. of Materials Science, Universit\`{a} di Milano-Bicocca, via R. Cozzi 53, I-20125 Milano, Italy}

\author{Michele Parrinello} 
\affiliation{Computational Science, DCHAB, ETH Zurich, USI Campus, via G. Buffi 13, CH-6900 Lugano, Switzerland}

\begin{abstract}
Owing to their small mass, hydrogen atoms exhibit strong quantum behavior even at room
temperature. Including these effects in first principles calculations is 
challenging, because of the huge computational effort required by 
conventional techniques.
Here we present the first ab-initio application of a recently-developed stochastic 
scheme, which allows to approximate nuclear quantum effects inexpensively. 
The proton momentum distribution of lithium imide, 
a material of interest for hydrogen storage, was experimentally 
measured by inelastic neutron scattering experiments
and compared with the outcome of quantum thermostatted 
ab initio dynamics. 
We obtain favorable agreement between theory and experiments
for this purely quantum mechanical property, thereby demonstrating
that it is possible to improve the modelling of complex hydrogen-containing
materials without additional computational effort.
\end{abstract}


\maketitle

Nuclear quantum effects play an important role in determining the properties 
of compounds containing light elements, hydrogen in particular. 
In order to assist the interpretation of experiments, accurate theoretical
modeling is highly desirable. Unfortunately, conventional techniques\cite{feyn-hibb65book,cepe95rmp}
can be orders of magnitude more expensive than methods which treat the nuclei
as classical particles. As a consequence, in ab initio simulations nuclear quantum effects
have seldom been included\cite{marx-parr96jcp,marx+99nat}.

A stochastic molecular dynamics framework based on generalized Langevin equations has
been recently devised. Among the many possible 
applications\cite{ceri+09prl,ceri+09prl2,ceri+10jctc}, it allows 
one to model to a good approximation nuclear 
quantum effects at negligible additional effort with respect to purely classical 
dynamics.
Preliminary tests based on empirical force fields demonstrated 
satisfactory agreement with path integral and experimental results\cite{ceri+09prl2,ceri+10jctc}.

An additional advantage of this approach is that not only atomic configurations,
but also the momentum reproduce the quantum distribution.  
On the contrary, computing the momentum distribution
within a path integral formalism\cite{cepe95rmp,morr-car08prl}
requires special techniques and, despite recent developments\cite{lin2010},
further increases the computational effort.
In the classical limit, the distribution of the momentum $\mathbf{p}$ of a particle is
Gaussian, $n(\mathbf{p})\propto \exp( -\mathbf{p}^2/2m k_B T)$, 
and depends only on the temperature $T$ and the particle's mass $m$. 
Conversely, in a quantum mechanical description $n(\mathbf{p})$ reflects
the local potential experienced by the particle. 
Deviation of the proton momentum distribution (PMD) from the classical 
one is a very sensitive probe of the quantum-mechanical behavior of 
hydrogen atoms. 
Experimental measurements of the PMD have been made feasible since the 
advent of spallation neutron sources. Indeed, the intense fluxes of 
neutrons in the 1-100 eV energy range provided by these facilities allows
one to study the short time ($10^{-16}$~s) dynamical properties of the proton 
in different hydrogen containing systems, as well as quantum fluids 
and solids\cite{andre+05advp}.

In this Letter we apply for the first time the ``quantum thermostat''
together with ab initio molecular dynamics, in order to model lithium imide.
Besides its technological significance as a material for
hydrogen storage\cite{chen+02nat,chen-zhu08mt}, $\mathrm{Li_2NH}$ is well suited as a benchmark.
In fact, the presence of libration modes of the $\mathrm{NH}$ bonds is likely
to introduce significant anharmonicities, which rule out the possibility of 
an accurate treatment by harmonic lattice dynamics. 

Theoretical results are compared
with the experimentally-determined PMD in lithium imide, which 
was obtained by means of Deep Inelastic Neutron Scattering (DINS) measurements 
on the VESUVIO spectrometer at the ISIS spallation neutron source 
(Rutherford Appleton Laboratory, United Kingdom)\cite{ISIS}, 
in the Resonance Detector (RD) configuration and using the 
Foil Cycling Technique\cite{FCT1,FCT2} that provides a narrow resolution
 suitable for line shape analysis on PMD.  
The high energy and wave vector transfers achievable with DINS allow one
to describe the scattering event within the framework of the impulse 
approximation (IA) with a very good degree of accuracy\cite{andre+05advp,IA}, 
so as to extract the proton momentum distribution directly from the experimental data. 
Actually the value of the wave vector transfer at the maximum of the proton recoil peak 
ranges from 35 \AA$^{-1}$ to 150 \AA$^{-1}$ for the complete set of detectors used in 
the present experiment. 
Following Refs.\cite{IAA,IAB}, one can easily evaluate the coefficient which multiplies 
the first term beyond the impulse approximation, namely the third derivative of 
the IA response function itself.
Using appropriate physical quantities for lithium imide, one finds 
(in the low temperature limit and for the aforementioned values of the wave vector 
transfer) that this "final state effect" coefficient lies in between 
3.67 \AA$^{-3}$ and 0.856 \AA$^{-3}$. 
This ensures that the impulse approximation is already reached during 
the reported neutron scattering measurement to any practical purpose.

The polycrystalline sample of Li$_2$NH was synthesized by thermal 
decomposition of commercial lithium amide (Sigma-Aldrich Inc., reagent grade) 
at $T= 623$~K and $p =10^{-3}$~Pa for 4~h,  
according to the reaction 
LiNH$_2$ $\rightarrow$ $\frac{1}{2}$ Li$_2$NH + $\frac{1}{2}$ NH$_3$ \cite{kojima,noritake}, 
in a furnace equipped with turbomolecular vacuum pump.  
X-ray powder diffraction measurements (CuK$_\alpha$ radiation) showed the sample 
to be  well crystallized and to contain a small quantity of Li$_2$O, which was already 
present in the pattern of the starting lithium amide. 
This minor contamination was reported also in the previous studies\cite{kojima,noritake};
on the other hand, no traces of LiOH or of other impurities were observed.

Being a powdered sample, DINS measures the spherically averaged PMD.
The detailed experimental procedures to extract 
the PMD from DINS data can be found in Refs.\cite{APa,APb,APc,APd} where data reduction and line shape 
analysis procedures are described in details.

\begin{figure}
\caption{\label{fig:suxcell} A representation of 
the initial configuration of the atoms in the supercell used for our
simulations\cite{mice+10nuovo}. The stable, tetrahedral clusters of interstitial Li atoms
are highlighted.
}
\includegraphics[width=1.0\columnwidth]{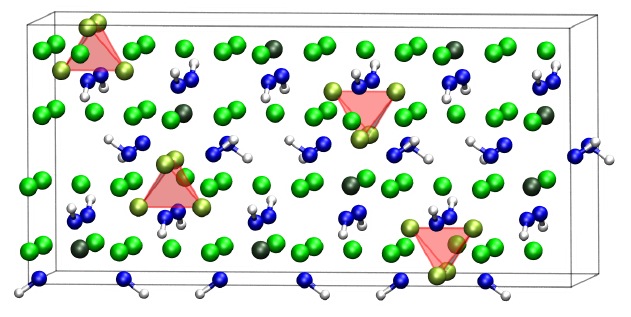}
\end{figure}

Simulations were performed using a supercell containing 
$192$-atoms~(Figure~\ref{fig:suxcell}), which were 
arranged according to the partially-disordered structure
which was recently proposed as a model of the low-temperature,  
$Fd\bar{3}m$ phase of $\mathrm{Li_2NH}$\cite{mice+10nuovo}. 
In this structure, Li atoms occupy the 
tetrahedral sites of the $fcc$ lattice of nitrogen atoms. Two
types of Li vacancies with different local symmetry are present.
One kind of vacancy is tetrahedrally coordinated
by N--H groups, while the second one is tetrahedrally coordinated
by Li interstitials.
These tetrahedral clusters of Li interstitials are found to stabilize the structure considerably, 
and are distributed in a disordered way, resulting in a excellent match 
with experimental diffraction data\cite{mice+10nuovo,balo+06jac}.
Moreover, they hinder the mobility of Li interstitials, which 
would be very high if the clusters were broken\cite{mice+10nuovo}.
Starting from this structure we performed Born-Oppenheimer 
molecular dynamics simulations within Density Functional Theory 
(DFT) with gradient corrected exchange and correlation functional\cite{perd+96prl} 
as implemented in the CPMD\cite{CPMD} package. Technical details
of the calculations are the same as those used in Ref.\cite{mice+10nuovo}.
Thermal averages were performed over 15~ps of trajectory, following 5~ps used
for equilibration at $T=300$~K.

Nuclear quantum effects were treated by means of a quantum thermostat, 
which is based on a bespoke generalized Langevin equation
containing correlated noise. This stochastic process is designed to mimic the 
quantum mechanical phase space distribution in the harmonic limit. 
The resulting non-equilibrium dynamics samples a stationary distribution which
is a good approximation of the quantum mechanical one in fairly anharmonic
systems as well. This result is achieved without requiring any 
preliminary information, except for an upper-bound estimate of the stiffest 
vibrational mode present. 
We used the set of parameters {\sffamily qt-50\_6}, which can be downloaded
from an on-line repository\cite{gle4md}. 
We refer the reader to Refs.\cite{ceri+10jctc,gle4md} for further details.

\begin{figure}
\caption{\label{fig:gr-comparison}(color online) Comparison between the intramolecular peak
of the N--H radial distribution function, as computed from molecular dynamics (MD)
and harmonic lattice dynamics (HLD)\cite{koha+92prb}, with and without considering nuclear quantum 
effects. Note that because of strong anharmonicities the classical HLD provides 
an unsatisfactory description of this system. A similar discrepancy is found 
when comparing the quantum thermostat results and HLD with Bose-Einstein occupations.
}
\includegraphics{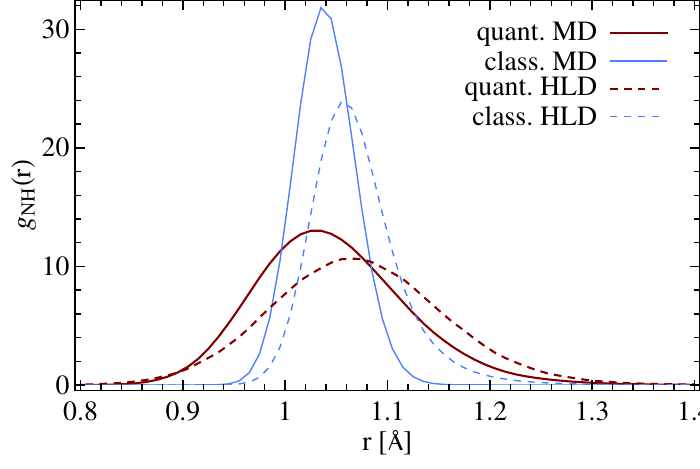}
\end{figure}

Before discussing the comparison with the experimental proton
momentum distribution, it is necessary to stress that the presence of 
anharmonic wagging modes of the imide groups makes a quasi-harmonic treatment
inappropriate. This is apparent in figure~\ref{fig:gr-comparison}, where  
we compare the radial distribution function of the N--H group as computed from 
molecular dynamics and from the harmonic normal modes. 
The same figure also highlights the importance of nuclear quantum effects,
which modify dramatically the typical fluctuations of the imide bonds.

While hydrogen and the stretching mode in particular exhibit the 
largest deviations from classical behavior,
lithium and nitrogen are also light nuclei and are therefore 
subject to nuclear quantum effects, albeit to a lesser extent. 
For instance, the average kinetic temperature computed during the 
quantum-thermostatted dynamics deviates from the classical value, and
is $415$~K for Li atoms, $410$~K for N (which is heavier but participates into
the stiff N--H stretching mode) and $858$~K for the protons.
These deviations illustrate the importance that nuclear quantum effects 
have in determining the properties of $\mathrm{Li_2NH}$, 
such as the temperature at which the transition between the low-temperature
and high-temperature phases occurs\cite{balo+06jac}.

\begin{figure}
\caption{\label{fig:pm3d} Three-dimensional proton momentum distribution
for the low temperature phase of lithium imide. 
Isosurfaces enclose 95\%, 90\%, 50\% and 10\% of the probability density. 
The arrangement of the hydrogens around a Li vacancy, aligned along the $\left<111\right>$ 
axes, is also reported, relative to the Cartesian reference.
}
\includegraphics[width=0.8\columnwidth]{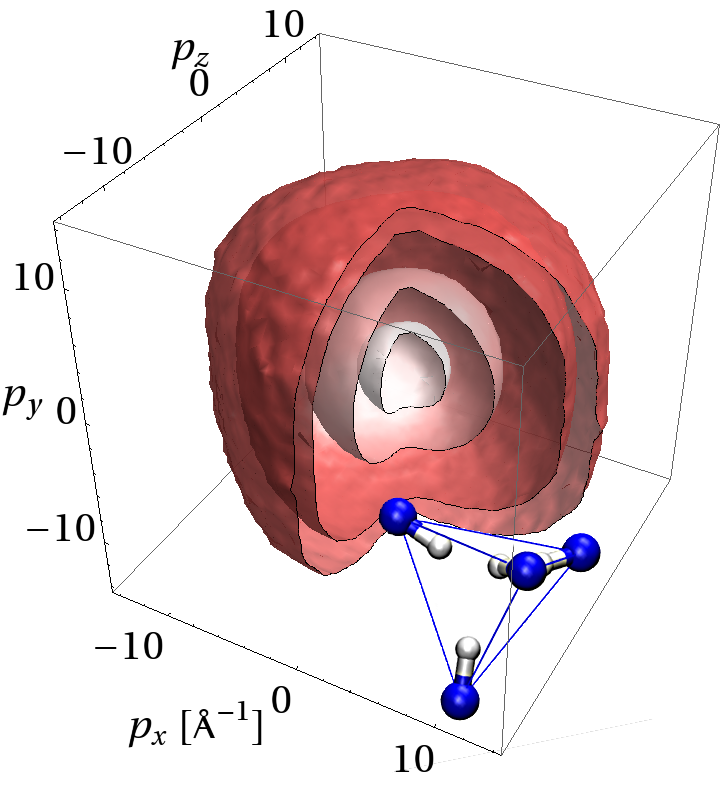}
\end{figure}

Computing the proton momentum distribution from the quantum-thermostatted dynamics
is straightforward. In fact, the thermostat has been designed to yield the correct
momentum and position distribution in the harmonic limit, and has proven to work also
in the anharmonic case. Thus we only need to collect 
the momentum histogram to obtain the three-dimensional
PMD, which is plotted in figure~\ref{fig:pm3d}. The anisotropy of $n(\mathbf{p})$, which is
a purely quantum mechanical effect, reflects the symmetry of the local environment of
protons inside the crystal, with the bonds aligned along $\left<111\right>$ directions of the 
nitrogen antifluorite sublattice, pointing towards 
Li vacancies\cite{hect-herb08jpcm,balo+06jac,mice+10nuovo}.
While we could not measure the directionally-resolved 
$n(\mathbf{p})$ because of the difficulties in obtaining
a single-crystal sample of appropriate dimensions, this result demonstrates
the ease by which this quantity can be accessed by our computational technique,
providing detailed information which can help to interpret experiments for
other systems\cite{KDP}. 

\begin{figure}
\caption{\label{fig:pmd}(color online) Comparison between the
spherically-averaged proton-momentum distribution 
expected for a classical system and from harmonic lattice dynamics at $T=300$~K, 
the PMD measured from a lithium imide sample and 
that computed from quantum-thermostatted molecular dynamics. }
\includegraphics{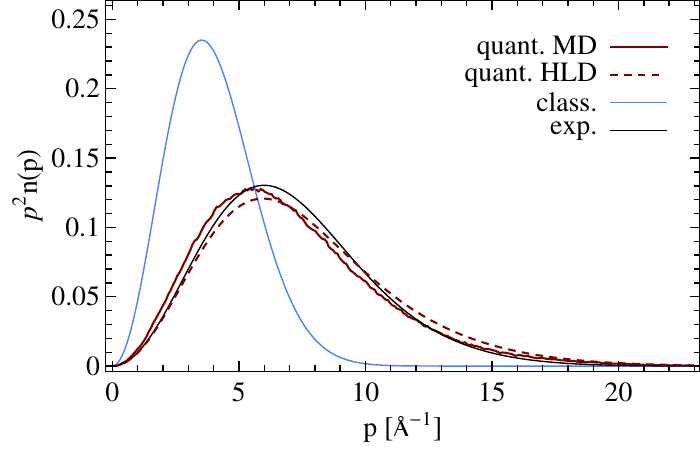}
\end{figure}

To compare with the powder sample experimental data, we 
spherically averaged the three dimensional PMD.
As it can be seen from figure~\ref{fig:pmd}, there is a satisfactory agreement 
between experiments and theory. In particular, we observe that 
quantum-thermostatted simulations match experimental data better than
the results from the quantum harmonic approximation.
To quantify the discrepancy with experiments, and to extract information relevant 
to the structural model of $\mathrm{Li_2NH}$,
we fitted the experimental and theoretical PMD's with a anisotropic Gaussian 
model.
We assumed a different spread in the directions parallel and perpendicular to the 
N--H bond\cite{reit+04bjp,garbuio}, resulting into 
\begin{equation}
n(\mathbf{p})\propto \exp\left(-\frac{p_z^2}{2\sigma^2_\parallel}-\frac{p_x^2+p_y^2}{2\sigma^2_\perp}\right),
\label{eq:pmd-model}
\end{equation}
which was then spherically averaged.

The resulting fit matches both curves very well, with values of adjusted $R^2$ greater 
than 0.999. The fit yields the parameters
$\sigma_\parallel =6.49$~\AA$^{-1}$ and $\sigma_\perp=3.15$~\AA$^{-1}$
for the theoretical PMD, and 
$\sigma_\parallel =5.77$~\AA$^{-1}$ and $\sigma_\perp=3.70$~\AA$^{-1}$
for the experimental one. 

The discrepancy is certainly larger than the experimental error bar,
however the anisotropy in the PMD is correctly captured, and the spread
in the direction parallel and perpendicular to the bond is qualitatively reproduced. 
This is indeed a remarkable result for an approximate model of nuclear quantum 
effect such as the one used in the present work, 
which can be applied with no computational overhead with respect to standard 
ab initio molecular dynamics.
While it is difficult to assess the uncertainty in our theoretical results,
a possible approach to gauge the error is to repeat simulations with a different
set of noise parameters. We did so using the parameters {\sffamily qt-20\_6}\cite{gle4md}, 
and obtained $\sigma_\parallel =6.31$~\AA$^{-1}$ and $\sigma_\perp=3.40$~\AA$^{-1}$. 
The comparison with the results obtained using {\sffamily qt-20\_6} 
sets a lower bound for the error at about $10$\%. 
Although the difference between the PMD obtained for different structures proposed
for $\mathrm{Li_2NH}$\cite{mice+10nuovo,balo+06jac,magy+06prb,muel-cede06prb} 
is smaller than our error bar above on the spherically-averaged distribution,
the difference in structure reflects in qualitatively-different 
three-dimensional $n(\mathbf{p})$.
However, for the reasons discussed in our previous work\cite{mice+10nuovo}, 
the other proposed structures have to be dismissed because of worse agreement with 
diffraction data. 

In conclusion, we have shown how a recently-developed method to
compute nuclear quantum effects can be used together with ab initio 
molecular dynamics to model complex materials containing light atoms.
We were able to use a large simulation cell, which was mandatory for
$\mathrm{Li_2NH}$ to reproduce the experimental structure of the 
low-temperature phase.
We obtained good agreement with the experimental
proton momentum distribution, a result of great significance given the
growing importance of inelastic neutron scattering experiments
as a sensitive probe of the local environment
in hydrogen-containing materials. 
Both experimental and theoretical data
are perfectly fitted by a model in which imide groups perform hindered 
librations.

The possibility of treating delicate nuclear quantum effects inexpensively,
albeit approximately, suggests that the quantum thermostat should be used
whenever light atoms are present, and a more accurate treatment by path
integral methods is unfeasible because of the excessive computational 
effort. Together with accurate first-principles calculations of the 
interatomic forces, this will shed light on the role of nuclear
quantum effects in condensed-phase systems. 
R. Senesi and J. Mayers  are gratefully acknowledged for suggestions 
during data reduction and analysis. We thank 
C. Andreani for useful discussions. 
One of the authors (AP) acknowledges the CNISM-CNR research program.

\end{document}